\documentclass{elsarta}
\begin{document}
\hspace*{3.5 in}CUQM-118\\
\hspace*{3.5 in}math-ph/0610008\\
\begin{frontmatter}
\title{Wave equation and dispersion relations for a compressible
rotating fluid}
\author{Jos\'e Mar\'\i n-Antu\~na$^\dagger$, Richard L. Hall$^\ddagger$, and Nasser Saad$^\ast$}
\address{${}^\dagger$Department of Theoretical Physics, Faculty of Physics,
University of Havana.} \ead{marin@fisica.uh.cu}

\address{${}^\ddagger$Department of Mathematics and Statistics, Concordia University,
1455 de Maisonneuve Boulevard West, Montr\'eal, Qu\'ebec, Canada H3G
1M8} \ead{rhall@mathstat.concordia.ca}
\address{${}^\ast$Department of Mathematics and Statistics,
University of Prince Edward Island, 550 University Avenue,
Charlottetown, Prince Edward Island, Canada C1A 4P3}
\ead{nsaad@upei.ca}

\begin{abstract}
A fundamental non-classical fourth-order partial differential
equation to describe small amplitude linear oscillations in a
rotating compressible fluid, is obtained. The dispersion relations
for such a fluid, and the different regions of the group and phase
velocity are analyzed.
\end{abstract}

\markboth{J. Mar\'\i n-Antu\~na, R. L. Hall and N. Saad}{Some
problems of propagation $\dots$}


\end{frontmatter}

\maketitle \markboth{J. Mar\'\i n-Antu\~na, R. L. Hall and N.
Saad}{Some problems of propagation $\dots$}
\section{Introduction} 
In this paper we study the dynamics of a rotating compressible
fluid, taking into account such factors as non-homogeneity, and the
presence of Coriolis forces due to the Earth rotation. The work is
devoted principally to the study of one of these aspects of the
problem, namely the influence of the rotation over the wave
propagation in the fluid.

Much work has been devoted to theoretical, mathematical, and
physical aspects of waves in compressible and incompressible fluids.
The literature of this subject is enormous, primarily because it
provides a basis for a wide range of applications and techniques,
including the solution and modeling of Geophysical situations, and
in the theory of stars. For example the papers
\cite{chas}-\cite{sot}, and books which discuss the fundamental
equations governing waves in such fluids \cite{jesus}. More
particularly, study of problems of wave propagation in a rotating
compressible fluid have been the subject of several earlier
publications \cite{mam} - \cite{gaam}. Some works have been devoted
to the general theory of equations in stratified and rotating
fluids~\cite{gaa} - \cite{gaaj}. Some monographs are devoted to this
type of problem~\cite{bga} - \cite{ll}. Nevertheless, with few
exceptions \cite{mas}, the systematic analtytic study of concrete
non-stationary problems of wave propagation in a rotating
compressible fluid have not been treated.

The present line of investigation has its origin in the seminal work
of S. L. Sobolev \cite{sob}. Starting from the hydrodynamic
equations we obtain in Sec.~2 a non-classical fourth-order partial
differential equation, which describes small amplitude linear
oscillations in such a fluid. In Sec.~3 we analyse the dispersion
relations for linear waves in the fluid, and discuss the different
numerical regions of the group and phase velocities of the
propagating waves, including their dependence on the direction of
the wave vector with respect to the rotation axis of the fluid.

The results obtained are potentially useful for the study of the
structure of the hydrophysical fluid fields in oceans and in the
Earth atmosphere.
\section{Basic equations} 
Consider an ideal compressible fluid that fills the whole space. We
assume that the fluid rotates with constant velocity $\alpha/2$
around a given axis. The fluid motion is refered to a system of
Cartesian coordinates $\vec{x}=(x_1,x_2,x_3)$ which rotates with the
fluid. The $Ox_3$ axis is directed along the rotation axis.

For such conditions, following ideas developed in \cite{bga} and
\cite{whi}, the system of hydrodynamic equations which describe the
fluid is given by the following:

\begin{enumerate}

\item The Euler equation:
\begin{equation}
\label{1} \frac{\partial \vec{v}}{\partial t} + (\vec{v} \cdot
\vec{\nabla}) \vec{v} + \frac{1}{\rho} \vec{\nabla} P - \vec{f} = 0,
\end{equation}
where $\vec{v}=(v_1,v_2,v_3)$ is the vector velocity of the fluid
particles, $P$ is the pressure, $\rho$ is the density of the fluid,
and $\vec{f}$ is the force per unit mass acting on the fluid.\medskip

\item The continuity equation

\begin{equation}
\label{2} \frac{d \rho}{d t} + \rho (\vec{\nabla} \cdot \vec{v}) =
0.
\end{equation}

\item The Thermodynamic state equation
\begin{equation}
\label{3} P = P(s,t),
\end{equation}
where $s$ is the entropy of the fluid.\medskip
\end{enumerate}

If we denote $\left(\frac{\partial P}{\partial \rho} \right)_s$ by
$c^2$, that is to say

\begin{equation}
\label{4} c^2 = \left( \frac{\partial P}{\partial \rho} \right)_s,
\end{equation}

then we have

\begin{equation}
\label{5} \frac{d P}{d t} = c^2 \frac{d \rho}{d t},
\end{equation}

where $c^2$ is a function of $P$ and $\rho$, and $c$ has the meaning
of the sound velocity in the fluid. Taking into account (\ref{5}),
Eq.~(\ref{2}) can be written

\begin{equation}
\label{6} \frac{1}{c^2} \frac{d P}{d t} + \rho \vec{\nabla} \cdot
\vec{v} = 0
\end{equation}

for iso-entropic motions in the fluid. In (\ref{1})
$\vec{f}$ is the Coriolis force due to the rotation of the fluid.
This means

\begin{equation}
\label{7} \vec{f} = -\vec{\alpha} \times \vec{v}.
\end{equation}

The corresponding Euler equation now becomes

\begin{equation}
\label{8} \frac{\partial \vec{v}}{\partial t} + (\vec{v} \cdot
\vec{\nabla}) \vec{v} + \frac{1}{\rho} \vec{\nabla} P + \vec{\alpha}
\times \vec{v} = 0.
\end{equation}

Eq. (\ref{8}) is nonlinear. Therefore, the exact theory of physical
processes in the fluid is nonlinear. In this paper, we study a
linear approximation, and, to this end, we consider small
perturbations in the fluid. We therefore linearize equation
(\ref{8}). The nonlinear term in (\ref{8}) is the inertial one,
$(\vec{v} \cdot \vec{\nabla}) \vec{v}$. We let $v_0$ be the velocity
amplitude, $\gamma$ the frequency of the wave, and $k$ the wave
number. Then, we have

\begin{equation}
\label{9} \varepsilon= \frac{(\vec{v} \cdot \vec{\nabla})
\vec{v}}{\frac{\partial \vec{v}}{\partial t}} = \frac{k
v_0^2}{\gamma v_0} = \frac{v_0}{v_{ph}},
\end{equation}

where $v_{ph} = \gamma/ k$ is the phase velocity of the wave.
$\varepsilon$ is a parameter of nonlinearity. We assume $\varepsilon
\ll 1$, that means that the velocity of the fluid particles on the
wave is much less than the phase velocity of the wave. Under these
conditions, we can eliminate the inertial term and obtain

\begin{equation}
\label{10} \frac{\partial \vec{v}}{\partial t} + \vec{\alpha} \times
\vec{v} + \frac{1}{\rho} \vec{\nabla} P = 0.
\end{equation}

This equation, along with (\ref{6}), comprises the system of
equations we shall study. We consider small displacements of the
pressure and of the density with respect to an equilibrium values:

$$
P=p_0+p, \rho = \rho_0+ \rho', p \ll p_0, \rho' \ll \rho_0.
$$

Under these conditions (\ref{10}) gives

$$
\frac{\partial \vec{v}}{\partial t} + \vec{\alpha} \times \vec{v} +
\frac{1}{\rho_0 + \rho'} \vec{\nabla} (p_0+ p) =0,
$$

and (\ref{6}) gives:

$$
\frac{1}{c^2} \frac{d(p_0+p)}{dt}+ (\rho_0+\rho') \vec{\nabla} \cdot
\vec{v}=0.
$$

By neglecting $\rho'$ in these expressions, we arrive at the
linearized system of equations used to describe the small-amplitude
motion in the fluid:

\begin{equation}
\label{11} \frac{\partial \vec{v}}{\partial t} + \vec{\alpha} \times
\vec{v} + \frac{1}{\rho_0} \vec{\nabla} p = 0,
\end{equation}

\begin{equation}
\label{12} \frac{1}{c^2} \frac{d p}{d t} + \rho_0 \vec{\nabla} \cdot
\vec{v} = 0,
\end{equation}

where $p$ is the dynamical displacement of the pressure from the
equilibrium position $p_0$. System (\ref{11}) - (\ref{12}) is a
4-equation system of first order, with 4-unknown variables, $p$,
$v_1$, $v_2$, and $v_3$, which describes the small-amplitude motion
(acoustic waves) in the ideal rotating fluid.

We will consider what we will call 2D motion in the fluid: by this
we mean variation of the dynamic pressure and of the velocity for
which
$$
\frac{\partial p}{\partial x_2}=0,
$$
and
$$
\frac{\partial \vec{v}}{\partial x_2}=0.
$$
From the physical and geometrical points of view, these kinds of motion are possible only in domains with infinite cylindrical shape, with the generatrices parallel to the $Ox_2$ axis. In what follows, we suppose that the main value of the density is given by $\rho_0=1$. Then, the system (\ref{11})-(\ref{12}) in the variables $v_1$, $v_2$, $v_3$, and $p$ becomes

\begin{equation}
\label{13} \frac{\partial v_1}{\partial t} - \alpha v_2 +
\frac{\partial p}{\partial x_1} = 0,
\end{equation}

\begin{equation}
\label{14} \frac{\partial v_2}{\partial t} + \alpha v_1 = 0,
\end{equation}

\begin{equation}
\label{15} \frac{\partial v_3}{\partial t} + \frac{\partial
p}{\partial x_3} = 0,
\end{equation}

\begin{equation}
\label{16} \frac{1}{c^2}\frac{\partial p}{\partial t} +
\frac{\partial v_1}{\partial x_1} + \frac{\partial v_3}{\partial
x_3} = 0.
\end{equation}

From the system (\ref{13}) - (\ref{16}) it is straightforward to
obtain the following equation

\begin{equation}
\label{17} L[u] = \frac{\partial^2}{\partial t^2} \left[
\frac{1}{c^2} \frac{\partial^2 u}{\partial t^2} - \nabla_2^2 u +
\frac{\alpha^2}{c^2} u \right] - \alpha^2 \frac{\partial^2
u}{\partial x_3^2} = 0,
\end{equation}

where $ \nabla_2^2 \equiv \partial^2/\partial x_1^2 +
\partial^2/\partial x_3^2$ is the 2D Laplacian. Eq.~(\ref{17}) is
satisfied by the pressure $p$ and also by the components of the
velocity $v_1$, $v_2$, and $v_3$. This is a non-classical fourth
order partial differential equation, which governs the waves in a
rotating compressible fuid.

From (\ref{17}) we can see that, when $\alpha = 0$ (which means that
the fluid does not rotate), the operator $L$ has the form $L=
\frac{\partial^2}{\partial t^2} \left[ \frac{1}{c^2}
\frac{\partial^2 u}{\partial t^2}- \nabla_2^2 u \right]$, and
therefore the solution must coincide with the solution of the 2D
wave equation. This allows us to conclude that, as a result of the
rotation of the fluid, some terms appear in the equation that take
into account the rotation. The corresponding solutions reflect this
fact by their dependence on the angular velocity $\alpha$. Moreover,
when $\alpha \rightarrow 0,$ the solutions of reduce to the
well-known solutions for wave propagation in a compressible medium.
\section{Dispersion relations}
In what follows we shall adopt units in which $c=1$. Thus

\begin{equation}
\label{e20} L[u] = \frac{\partial^2}{\partial t^2} \left[
\frac{\partial^2 u}{\partial t^2} - \nabla_2^2 u + \alpha^2 u
\right] - \alpha^2 \frac{\partial^2 u}{\partial x_3^2} = 0.
\end{equation}

We consider a solution of the form

\begin{equation}
\label{e21} u=u_0 e^{i(\vec{k} \cdot \vec{x} + \gamma t)},
\end{equation}

where $\vec{k}=(k_1,k_3)$ is the wave vector, $\vec{x}=(x_1,x_3)$,
and $\gamma$ is the wave frequency. We obtain:

\begin{equation}
\label{e22} \gamma^2(\gamma^2- \alpha^2) - \gamma^2 k^2 + \alpha^2
k^2 \cos^2 \theta = 0,
\end{equation}

where $\theta$ is the angle between wave vector $\vec{k}$ and the
$Ox_3$ axis. Eq.~(\ref{e22}) is the dispersion relation for plane
waves in a rotating compressible fluid. We have:

\begin{equation}
\label{e23} k= |\gamma| \sqrt{\frac{\alpha^2 - \gamma^2}{\alpha^2
\cos^2 \theta - \gamma^2}}.
\end{equation}

We can conclude that there exists a fundamental difference between
the propagation of plane harmonic waves in a rotating fluid and
those in a fluid at rest. From (\ref{e23}) with $\alpha \rightarrow
0$  we obtain $k=|\gamma|$, that is to say, the well-known relation
for plane waves in a fluid at rest. We note the dependence of $k$ on
the value of angle $\theta$ (the direction of the wave vector $k$).
Two special cases are qualitatively different. In the case in which
$\vec{k}$ is parallel to the rotating axis, (\ref{e23}) gives
$k=|\gamma|$, which means that in the $Ox_3$ direction plane waves
with any frequency $\gamma$ propagate, including step-like waves
($\gamma=0$). When we have a wave vector $\vec{k}$ with direction
oblique to the rotation axis ($0< \theta < \pi/2$), we infer that
waves propagate only for certain values of $\gamma$: there is a
forbidden zone for the frequency $\gamma$, namely $\alpha \cos
\theta < \gamma < \alpha$. In the case where $\vec{k}$ is
perpendicular to the rotating axis ($\theta = \pi/2$) we have that
only waves with frequency $\gamma > \alpha$ propagate.

By differentiating the dispersion relation (\ref{e22}) we obtain the
group velocity of the waves, thus

\begin{equation}
\label{e24} \vec{v}_g = \frac{\gamma}{2 \gamma^2 - \alpha^2 - k^2}
\left (k_1, \frac{\gamma^2- \alpha^2}{\gamma^2}k_3 \right ).
\end{equation}

Meanwhile, we know that the phase velocity of the waves is

\begin{equation}
\label{e25} \vec{c} \equiv \vec{v}_{ph} = \frac{\gamma}{k}
\frac{\vec{k}}{k} = \frac{\gamma \vec{k}}{k^2}.
\end{equation}

From (\ref{e24}) we conclude the following:

\begin{enumerate}

\item For the fluid at rest ($\alpha=0$) and, hence, $k^2=\gamma^2$. We obtain

\begin{equation}
\label{e26} \vec{v}_{\rm g} = \frac{1}{\gamma} \vec{k} \equiv
\vec{v}_{\rm ph}.
\end{equation}

Thus the group velocity is equal to the phase velocity; there is no
dispersion, and the energy propagates in the direction $\vec{k}$.

\item In general, for $\alpha \neq 0$, the direction of the group velocity $\vec{v}_{\rm g}$ (that is to say, the direction of propagation of the energy \cite{whi}) does not coincide with the direction of the vector $\vec{k}$, and wave dispersion takes place.

\end{enumerate}

When the direction of $\vec{k}$ coincides with the rotating axis
$Ox_3$, i.e. when $\theta=0$, for the group velocity we have

\begin{equation}
\label{e27} \vec{v}_{\rm g} = \frac{\gamma}{2 \gamma^2 -\alpha^2 -
\gamma^2} \left ( 0, \frac{\gamma^2 -\alpha^2}{\gamma^2}k \right ) =
\frac{1}{\gamma} \vec{k} = \vec{v}_{\rm ph}
\end{equation}

since $k_1=0$, $k_3=k$, and $k^2=\gamma^2$ in this case. Hence, the
waves again propagate without dispersion.

If $\theta= \pi/2$, and under the propagating conditions
$\gamma>\alpha$, we have from (\ref{e23})
$k=\sqrt{\gamma^2-\alpha^2}$. Therefore, for the group velocity, we
have

\begin{equation}
\label{e28} \vec{v}_{\rm g} = \frac{\gamma}{2 \gamma^2 -\alpha^2
-\gamma^2+\alpha^2} (k,0 ) = \frac{1}{\gamma} \vec{k}.
\end{equation}

Meanwhile from (\ref{e24}),

\begin{equation}
\label{e29} \vec{v}_{\rm ph} = \frac{\gamma}{\gamma^2-\alpha^2} \vec{k}.
\end{equation}

This means that $v_g<v_{\rm ph}$ and, therefore, we have in this case a
normal dispersion of the waves. This is in complete agreement with
the fact that, for $\theta= \pi/2$ and $k=
\sqrt{\gamma^2-\alpha^2},$ the waves disperse. For $|\gamma|<
\alpha$ there is a damping of waves travelling perpendicular to the
$Ox_3$ axis direction.

In the general case we can see, using (\ref{e22}), that the expresion
$2 \gamma^2 - \alpha^2 - k^2$ in the denominator of (\ref{e24}) is
always a positive number. From (\ref{e24}) it follows that the group-velocity vectors, corresponding to all possible plane waves
starting from the same begining, are inside the limits of a
characteristic cone

$$
|x_3|> \frac{\sqrt{\alpha^2 - \gamma^2}}{\gamma} |x_1|.
$$

The angle between the group vector and the rotating axis $Ox_3$ is

$$
\theta_{\rm gr} = \arctan \left[ \frac{\gamma^2}{\gamma^2 - \alpha^2}
\frac{k_1}{k_3} \right].
$$

\section{Conclusion} 
The general behaviour of acoustic waves in a rotating compressible
fluid can be described in a simple formulation using a fourth order
non-classical differential equation, which reduces to the classical
one when the fluid is at rest (no rotation). The analysis of the
dispersion relation exhibits the existence of a forbidden zone for
the frequencies of the propagating wave. We have also seen that, in
the presence of the non homogeneity of the space in which the waves
propagate (owing to the rotation of the fluid), the dispersion of
the waves takes place when the wave vector is not in the same
direction as the rotating axis. In future work we shall study
fundamental solutions of the basic equation (\ref{17}) for several
interesting problems involving the diffraction of waves at walls
within a rotating fluid.

\section*{Acknowledgments}
\medskip
\noindent Partial financial support of this work under Grant Nos.
GP3438 and GP249507 from the Natural Sciences and Engineering
Research Council of Canada is gratefully acknowledged by two of us
(respectively [RLH] and [NS]). One of us [JMA] acknowledges the
hospitality of the Department of Mathematics and Statistics of
Concordia University, where some of this work was carried out.

\end{document}